\documentclass{article}
\usepackage{amsthm, amssymb, amsmath}
\newtheorem{theorem}{Theorem}[section]

\newtheorem{remark}{Remark}

\newtheorem{corollary}[theorem]{Corollary}
\newenvironment{them}{\textbf{Theorem}\it}{}
\newtheorem{definition}[theorem]{Definition}
\long\def\symbolfootnote[#1]#2{\begingroup%
\def\thefootnote{\fnsymbol{footnote}}\footnote[#1]{#2}\endgroup}
\huge
\begin{document}
\title{On Soliton Dynamics in Nonlinear Schr\"odinger Equations\thanks{This paper is part of the first author's Ph.D
thesis.}}\date{}
\author{Zhou Gang\thanks{Supported by NSERC under Grant NA7901 and NSF under
Grant DMS-0400526.} \ I.M. Sigal$^{\dag}$}\maketitle
\centerline{\small{Department of Mathematics, University of Toronto,
Toronto, Canada}}
 \setlength{\leftmargin}{.1in}
\setlength{\rightmargin}{.1in} \setlength{\leftmargin}{.1in}
\setlength{\rightmargin}{.1in}
\section*{Abstract}
In this paper we announce the result of asymptotic dynamics of
solitons of nonlinear Schr\"odinger equations with external
potentials. To each local minima of the potential there is a soliton
centered around it. Under some conditions on the nonlinearity, the
potential and the datum, we prove that the solution can be
decomposed into two parts: the soliton and the term dissipating to
infinity.
\section{Introduction}
\textbf{Problem.} In this paper we study dynamics of solitons in the
generalized nonlinear Schr\"odinger equation (NLS) in dimension
$d\neq 2$ with an external potential $V_{h}:\
\mathbb{R}^{n}\rightarrow \mathbb{R}$,
\begin{equation}\label{NLS}
i\frac{\partial\psi}{\partial t}=-\Delta
\psi+V_{h}\psi-f(|\psi|^{2})\psi.
\end{equation}
Here $h>0$ is a small parameter giving the length scale of the
external potential in relation to the length scale of the $V_{h}=0$
solitons (see below), $\Delta$ is the Laplace operator and $f(s)$ is
a nonlinearity to be specified later. We normalize $f(0)=0$. Such
equations arise in the theory of Bose-Einstein condensation
\footnote[1]{In this case Equation (~\ref{NLS}) is called the
Gross-Pitaevskii equation.}, nonlinear optics, theory of water waves
\footnote[2]{In these two areas $V_{h}$ arises if one takes into
account impurities and/or variations in geometry of the medium and
is, in general, time-dependent.}and in other areas.

To fix ideas we assume the potentials to be of the form
$V_{h}(x):=V(hx)$ with $V$ smooth and decaying at $\infty.$ Thus
for $h=0,$ Equation (~\ref{NLS}) becomes the standard generalized
nonlinear Schr\"odinger equation (gNLS)
\begin{equation}\label{gNLS}
i\frac{\partial \psi}{\partial t}=-\Delta
\psi+\mu\psi-f(|\psi|^{2})\psi,
\end{equation} where $\mu=V(0).$
For a certain class of nonlinearities, $f(|\psi|^{2})$ (see Section
~\ref{ExSt}), there is an interval $\mathcal{I}_{0}\subset
\mathbb{R}$ such that for any $\lambda\in \mathcal{I}_{0}$ Equation
(~\ref{gNLS}) has solutions of the form
$e^{i(\lambda-\mu)t}\phi_{0}^{\lambda}(x)$ where
$\phi_{0}^{\lambda}\in \mathcal{H}_{2}(\mathbb{R}^{n})$ and
$\phi_{0}^{\lambda}>0.$ Such solutions (in general without the
restriction $\phi_{0}^{\lambda}>0$) are called the \textit{solitary
waves} or \textit{solitons} or, to emphasize the property
$\phi_{0}^{\lambda}>0,$ the \textit{ground states}. For brevity we
will use the term \textit{soliton} applying it also to the function
$\phi_{0}^{\lambda}$ without the phase factor $e^{i(\lambda-\mu)t}.$

Equation (~\ref{gNLS}) is translationally and gauge invariant. Hence
if $e^{i(\lambda-\mu)t}\phi_{0}^{\lambda}(x)$ is a solution for
Equation (~\ref{gNLS}), then so is $$
e^{i(\lambda-\mu)t}e^{i\alpha}\phi_{0}^{\lambda}(x+a),\ \text{for
any}\ a\in \mathbb{R}^{n},\text{and}\ \alpha\in [0,2\pi).$$ This
situation changes dramatically when the potential $V_{h}$ is turned
on. In general, as was shown in ~\cite{Floer,Oh1} out of the
$(n+2)$-parameter family
$e^{i(\lambda-\mu)t}e^{i\alpha}\phi_{0}^{\lambda}(x+a)$ only a
discrete set of two-parameter families of solutions to Equation
(~\ref{NLS}) bifurcate: $e^{i\lambda
t}e^{i\alpha}\phi^{\lambda}(x),$ $\alpha\in [0,2\pi)$ and
$\lambda\in \mathcal{I}$ for some $\mathcal{I}\subseteq
\mathcal{I}_{0}$, with $\phi^{\lambda}\equiv \phi_{h}^{\lambda}\in
\mathcal{H}_{2}(\mathbb{R}^{n})$ and $\phi^{\lambda}>0$. Each such
family centers near a different critical point of the potential
$V_{h}(x).$ It was shown in ~\cite{Oh2} that the solutions
corresponding to minima of $V_{h}(x)$ are orbitally (Lyapunov)
stable and to maxima, orbitally unstable. We call the solitary wave
solutions described above which correspond to the minima of
$V_{h}(x)$ \textit{trapped solitons} or just \textit{solitons} of
Equation (~\ref{NLS}) omitting the last qualifier if it is clear
which equation we are dealing with.

\textbf{Results}. In this note we describe results of
~\cite{GS1,GS2} that the trapped solitons of Equation (~\ref{NLS})
are asymptotically stable. The latter property means that if an
initial condition of (~\ref{NLS}) is sufficiently close to a trapped
soliton then the solution converges (relaxes),
$$\psi(x,t)-e^{i\gamma(t)}\phi^{\lambda_{\infty}}(x)\rightarrow 0,$$
in some weighted $\mathcal{L}^{2}$ space to, in general, another
trapped soliton of the same two-parameter family. We also find
effective equations for the soliton center and other parameters. In
this paper we prove this result under the additional assumption that
if $d>2$ then the potential is spherically symmetric and that the
initial condition symmetric with respect to permutations of the
coordinates. In this case the soliton relaxes to the ground state
along the radial direction. This limits the number of technical
difficulties we have to deal with. We expect that our techniques
extend to the general case when the soliton spirals toward its
equilibrium.

In fact, ~\cite{GS1,GS2} prove a result more general than asymptotic
stability of trapped solitons. Namely, we show that if an initial
condition is close (in the weighted norm
$\|u\|_{\nu,1}:=\|(1+|x|^{2})^{\frac{\nu}{2}}u\|_{\mathcal{H}^{1}}$
for sufficiently large $\nu$) to the soliton
$e^{i\gamma_{0}}\phi^{\lambda_{0}},$ with $\gamma_{0}\in \mathbb{R}$
and $\lambda_{0}\in \mathcal{I}$ ($\mathcal{I}$ as above), then the
solution, $\psi(t),$ of Equation (~\ref{NLS}) can be written as
\begin{equation}\label{decom1}
\psi(x,t)=e^{i\gamma(t)}\big(e^{ip(t)\cdot
x}\phi^{\lambda(t)}(x-a(t))+R(x,t)\big),
\end{equation}
where
$\|R(t)\|_{-\nu,1}\rightarrow 0$, $\lambda(t)\rightarrow
\lambda_{\infty}$ for some $\lambda_{\infty}$ as $t\rightarrow
\infty$ and the soliton center $a(t)$ and momentum $p(t)$ evolve
according to an effective equations of motion close to Newton's
equation in the potential $h^2V(a)$.

We observe that (~\ref{NLS}) is a Hamiltonian system with
conserved energy (see Section ~\ref{HaGWP}) and, though orbital
(Lyapunov) stability is expected, the asymptotic stability is a
subtle matter. To have asymptotic stability the system should be
able to dispose of excess of its energy, in our case, by radiating
it to infinity. The infinite dimensionality of a Hamiltonian
system in question plays a crucial role here. This phenomenon as
well as a general class of classical and quantum relaxation
problems was pointed out by J. Fr\"ohlich and T. Spencer [Private
Communication].

We also mention that because of slow time-decay of the linearized
propagator, the low dimensions $d=1,2$ are harder to handle than the
higher dimensions, $d>2.$

\textbf{Previous results}. We refer to ~\cite{GS1} for a detailed
review of the related literature. Here we only mention results of
~\cite{Cu,Buslaev,BP2,BuSu,SW1,SW2,SW3,TY1,TY2,TY3} which deal with
a similar problem.  Like our work, ~\cite{SW1,SW2,SW3,TY1,TY2,TY3}
study the ground state of the NLS with a potential. However, these
papers deal with the near-linear regime in which the nonlinear
ground state is a bifurcation of the ground state for the
corresponding Schr\"{o}dinger operator $-\Delta+V(x).$ The present
paper covers highly nonlinear regime in which the ground state is
produced by the nonlinearity (our analysis simplifies considerably
in the near-linear case). Now, papers ~\cite{Cu,Buslaev,BP2,BuSu}
consider the NLS without a potential so the corresponding solitons,
which were described above, are affected only by a perturbation of
the initial conditions which disperses with time leaving them free.
While in our case they, in addition, are under the influence of the
potential and they relax to an equilibrium state near a local
minimum of the potential.

\textbf{Open problems}. We formulate some open problems:

(1) Extend the results of this present paper to more general initial
conditions and to more general, probably time-dependent, potentials.

(2) Link the results of this paper with the results of ~\cite{FGJS}
on the long time dynamics of solitons.

A natural place to start here is spherically symmetric potentials
but general initial conditions. Note that for certain time-dependent
potentials the solitons will never settle in the ground state.

\textbf{Notation}. As customary we often denote derivatives by
subindices as in
$\phi^{\lambda}_{\lambda}=\frac{\partial}{\partial\lambda}\phi^{\lambda}$
for $\phi^{\lambda}=\phi^{\lambda}(x).$ However, the subindex $h$
signifies always the dependence on the parameter $h$ and not the
derivatives in $h.$ The Sobolev and $L^{2}$ spaces are denoted by
$\mathcal{H}^{k}$ and $\mathcal{L}^{2}$ respectively.

\textbf{Acknowledgment}. We are grateful to  J. Colliander, S.
Cuccagna, S. Dejak, J. Fr\"ohlich, Z. Hu, W. Schlag, A. Soffer, G.
Zhang, V. Vougalter and, especially, V.S. Buslaev for fruitful
discussions. This paper is part of the first author's Ph.D thesis
requirement.

\section{Hamiltonian Structure and GWP}\label{HaGWP}
Equation (~\ref{NLS}) is a Hamiltonian system on Sobolev space
$\mathcal{H}^{1}(\mathbb{R},\mathbb{C})$ viewed as a real space
$\mathcal{H}^{1}(\mathbb{R},\mathbb{R})\oplus
\mathcal{H}^{1}(\mathbb{R},\mathbb{R})$ with the inner product
$(\psi,\phi)=Re\int_{\mathbb{R}}\bar{\psi}\phi$ and with the
simpletic form
$\omega(\psi,\phi)=Im\int_{\mathbb{R}}\bar{\psi}\phi.$ The
Hamiltonian functional is: $$H(\psi):=\int
[\frac{1}{2}(|\psi_{x}|^{2}+V_{h}|\psi|^{2})-F(|\psi|^{2})],$$ where
$F(u):=\frac{1}{2}\int_{0}^{u}f(\xi)d\xi.$

Equation (~\ref{NLS}) has the time-translational and gauge
symmetries which imply the following conservation laws: for any
$t\geq 0,$ we have
\begin{enumerate}
 \item[(CE)] conservation of energy: $H(\psi(t))=H(\psi(0));$
 \item[(CP)]
 conservation of the number of particles: $N(\psi(t))=N(\psi(0)),$ where $N(\psi):=\int
 |\psi|^{2}.$
\end{enumerate}
To address the global well-posedness of (~\ref{NLS}) we need the
following condition on the nonlinearity $f$. Below, $s_{+}=s$ if
$s>0$ and $=0$ if $s\leq  0$.
\begin{enumerate}
 \item[(fA)] The nonlinearity $f$ satisfies the
 estimate $|f^{'}(\xi)|\leq c(1+|\xi|^{\alpha-1})$ for some $\alpha\in
 [0,\frac{2}{(d-2)_{+}})$  and $|f(\xi)|\leq c(1+|\xi|^{\beta})$ for some $\beta\in[0,\frac{2}{d}).$
\end{enumerate}

The following result can be found in ~\cite{Cazenave}.\\
\begin{them}
Assume that the nonlinearity $f$ satisfies the condition (fA), and
that the potential $V$ is bounded. Then Equation (~\ref{NLS}) is
globally well posed in $\mathcal{H}^{1}$, i.e. the Cauchy problem
for Equation (~\ref{NLS}) with a datum $\psi(0)\in \mathcal{H}^{1}$
has a unique solution $\psi(t)$ in the space $\mathcal{H}^{1}$ and
this solution depends continuously on $\psi(0)$. Moreover $\psi(t)$
satisfies the conservation laws (CE) and (CP).
\end{them}
\section{Existence and Orbital Stability of Solitons}\label{ExSt}
In this section we review the question of existence of the solitons
(ground states) for Equation (~\ref{NLS}). Assume the nonlinearity
$f:\mathbb{R}\rightarrow \mathbb{R}$ is smooth and satisfies
\begin{enumerate}
 \item[(fB)] There exists an interval $\mathcal{I}_{0}\in \mathbb{R}^{+}$
s.t. for any $\lambda\in\mathcal{I}_{0}$, $-\infty\leq
\displaystyle\overline{\lim}_{s\rightarrow +
 \infty}\frac{f(s)}{s^{\frac{2}{d-2}}}\leq 0$
and $\frac{1}{\xi}\int_{0}^{\xi}f(s)d
 s>\lambda$ for some constant $\xi$, for $d>2$;
and
 $$U(\phi,\lambda):=-\lambda\phi^{2}+\int_{0}^{\phi^{2}}f(\xi)d\xi$$
 has a smallest positive root $\phi_{0}(\lambda)$ such that
 $U_{\phi}(\phi_{0}(\lambda),\lambda)\not=0$, for $d=1$.
 \end{enumerate}

It is shown in ~\cite{BL, Str} that under Condition (fB) there
exists a spherical symmetric positive solution $\phi^{\lambda}$ to
the equation
\begin{equation}\label{soliton}
-\Delta\phi^{\lambda}+\lambda\phi^{\lambda}-f((\phi^{\lambda})^{2})\phi^{\lambda}=0.
\end{equation}
\begin{remark}
Existence of soliton functions $\phi^{\lambda}$ for $d=2$ is proved
in ~\cite{Str} under different conditions on $f$.
\end{remark}
When the potential $V$ is present, then some of the solitons above
bifurcate into solitons for Equation (~\ref{NLS}). Namely, let, in
addition, $f$ satisfy the condition $|f^{'}(\xi)|\leq
c(1+|\xi|^{p}),$ for some $p< \infty$, and $V$ satisfy the condition
\begin{enumerate}
\item[(VA)] $V$ is smooth and $0$ is a non-degenerate local
minimum of $V$.
\end{enumerate}
Then, similarly as in ~\cite{Floer,Oh1} one can show that if $h$ is
sufficiently small, then for any $\lambda\in \mathcal{I}_{0V}$,
where
$$\mathcal{I}_{0V}:=\{\lambda|\lambda>-\displaystyle\inf_{x\in\mathbb{R}}\{V(x)\}\}\cap\{\lambda|\lambda+V(0)\in
\mathcal{I}_{0}\},$$ there exists a unique soliton
$\phi^{\lambda}\equiv\phi_{h}^{\lambda}$ (i. e. $\phi^{\lambda}\in
\mathcal{H}_{2}(\mathbb{R})$ and $\phi^{\lambda}>0$) satisfying the
equation
$$-\Delta \phi^{\lambda}+(\lambda+V_{h})\phi^{\lambda}-
f((\phi^{\lambda})^{2})\phi^{\lambda}=0$$ and the estimate
$\phi^{\lambda}=\phi^{\lambda+V(0)}_{0}+O(h^{3/2})$ where
$\phi_{0}^{\lambda}$ is the soliton of Equation (~\ref{soliton}).

Let $\delta(\lambda):=\|\phi^{\lambda}\|_{2}^{2}$. It is shown in
~\cite{GSS1} that the soliton $\phi^{\lambda}$ is a minimizer of the
energy functional $H(\psi)$ for a fixed number of particles
$N(\psi)=constant$ if and only if
$\delta^{'}(\lambda)>0.$
Moreover, it is shown in ~\cite{We2,GSS1} that under the latter
condition the solitary wave $\phi^{\lambda}e^{i\lambda t}$ is
orbitally stable.
Under more restrictive conditions (see ~\cite{GSS1}) on $f$ one can
show that the open set
%
\begin{equation}
\mathcal{I}:=\{\lambda\in \mathcal{I}_{0V}:\delta'(\lambda)>0\}
\end{equation}
is non-empty. Instead of formulating these conditions we assume in
what follows that the open set $\mathcal{I}$ is non-empty and
$\lambda\in \mathcal{I}$.

Using the equation for $\phi^{\lambda}$ one can show that if the
potential $V$ is redially symmetric then there exist constants $c,\
\delta>0$ such that
$|\phi^{\lambda}(x)|\leq ce^{-\delta|x|}\ \text{and}\
|\frac{d}{d\lambda}\phi^{\lambda}|\leq ce^{-\delta|x|},$
and similarly for the derivatives of $\phi^{\lambda}$ and
$\frac{d}{d\lambda}\phi^{\lambda}$.

\section{Linearized Equation and Resonances}\label{subslinear} We
rewrite Equation (~\ref{NLS}) as $\frac{d\psi}{dt}=G(\psi)$ where
the nonlinear map $G(\psi)$ is defined by
$G(\psi)=-i(-\Delta+\lambda+V_{h})\psi+if(|\psi|^{2})\psi.$
Then the linearization of Equation (~\ref{NLS}) can be written as
$\frac{\partial\chi}{\partial t}=\partial G(\phi^{\lambda})\chi$
where $\partial G(\phi^{\lambda})$ is the Fr\'echet derivative of
$G(\psi)$ at $\phi$. It is computed to be
\begin{equation}\label{defineoperator}
\partial
G(\phi^{\lambda})\chi=-i(-\Delta+\lambda+V_{h})\chi+if((\phi^{\lambda})^{2})\chi+2if^{'}((\phi^{\lambda})^{2})(\phi^{\lambda})^{2}Re\chi.
\end{equation}
This is a real linear but not complex linear operator. To convert
it to a linear operator we pass from complex functions to real
vector-functions $\chi\longleftrightarrow \vec{\chi}=\left(
\begin{array}{lll}
\chi_{1}\\
\chi_{2}
\end{array}
\right),
$ where $\chi_{1}=Re\chi$ and $\chi_{2}=Im\chi.$ Then
$\partial G(\phi^{\lambda})\chi\longleftrightarrow
L(\lambda)\vec{\chi}$ where the operator $L(\lambda)$ is given by
\begin{equation}\label{operaL}
L(\lambda) :=  \left(
\begin{array}{lll}
0&L_{-}(\lambda)\\
-L_{+}(\lambda)&0
\end{array}
\right),
\end{equation}
with
$L_{-}(\lambda):=-\Delta+V_{h}+\lambda-f((\phi^{\lambda})^{2}),$
and
$L_{+}(\lambda):=-\Delta+V_{h}+\lambda-f((\phi^{\lambda})^{2})-2f^{'}((\phi^{\lambda})^{2})
(\phi^{\lambda})^{2}.$
The operator $L(\lambda)$ is extended to the complex space
$\mathcal{H}^{2}(\mathbb{R},\mathbb{C})\oplus
\mathcal{H}^{2}(\mathbb{R},\mathbb{C}).$ If the potential $V_{h}$
in Equation (~\ref{NLS}) decays at $\infty$, then by a general
result
$$\sigma_{ess}(L(\lambda))=(-i\infty,-i\lambda]\cap
[i\lambda,i\infty).$$

The eigenfunctions of $L(\lambda)$ are described in the following
theorem (cf ~\cite{GS1}, ~\cite{GS2}).

\begin{theorem}\label{mainpo}
Let $V$ satisfy Condition (VA) and $|h|$ be sufficiently small. Then
the operator $L(\lambda)$ has at least $2d +2$ eigenvectors and
associated eigenvectors with eigenvalues near zero: two-dimensional
space with the eigenvalue 0 and a $2d $-dimensional space with
non-zero imaginary eigenvalues $\pm i\epsilon_j(\lambda),$
$$\epsilon_j(\lambda):=h\sqrt{2e_j} +o(h),$$ where $e_j$ are
eigenvalues of the Hessian matrix of $V$ at value $x=0,$
$V^{''}(0)$. The corresponding eigenfunctions
$\left(\begin{array}{lll}
\xi_j\\
\pm i\eta_j
\end{array}
\right)$ are related by complex conjugation and satisfy
$$
\xi_j= \sqrt{2}\displaystyle \partial_{x_{k}}\phi_{0}^{\lambda}
 +o(h)\
\text{and}\ \eta_j= -h \sqrt{e_j}\displaystyle
x_{j}\phi^{\lambda}_{0} +o(h),$$ and $\xi_{i}$ and $\eta_{j}$ are
real.
\end{theorem}

\begin{remark}
The zero eigenvector $\left(
\begin{array}{lll}
0\\
\phi^{\lambda}
\end{array}
\right)$ and the associated zero eigenvector $\left(
\begin{array}{lll}
\partial_{\lambda}\phi^{\lambda}\\
0
\end{array}
\right)$ are related to the gauge symmetry $\psi(x,t)\rightarrow
e^{i\alpha}\psi(x,t)$ of the original equation and the $2d$
eigenvectors $\left(\begin{array}{lll}
\xi_j\\
\pm i\eta_j
\end{array}
\right)$ with $O(h)$ eigenvalues originate from the zero
eigenvectors $\left(
\begin{array}{lll}
\partial_{x_{k}}\phi_{0}^{\lambda}\\
0
\end{array}
\right), k=1,2,\cdot\cdot\cdot,d,$ and the associated zero
eigenvectors $\left(
\begin{array}{lll}
0\\
x_{k}\phi_{0}^{\lambda}
\end{array}
\right),\ k=1,2,\cdot\cdot\cdot,d,$ of the $V=0$ equation due to the
translational symmetry and to the boost transformation
$\psi(x,t)\rightarrow e^{ib\cdot x}\psi(x,t)$ (coming from the
Galilean symmetry), respectively.
\end{remark}

For  $d\geq 2$ we will be interested in permutationally symmetric
functions, $g\in \mathcal{L}^{2}(\mathbb{R}^{d})$, characterized
as
$$g(x)=g(\sigma x)\ \text{for any}\ \sigma\in S_{d}$$ with $S_{d}$ being the group of
permutation of $d$ indices and $\sigma
(x_{1},x_{2},\cdot\cdot\cdot,x_{d}):=(x_{\sigma(1)},x_{\sigma(2)},
\cdot\cdot\cdot,x_{\sigma(d)}).$
\begin{remark}\label{remark2}
For any function of the form $e^{ip\cdot x}\phi(|x-a|)$ with $a
\parallel p$, there exists a rotation $\tau$ such that
the function $e^{ip\cdot \tau
{x}}\phi(|\tau{x}-a|)=e^{i\tau^{-1}p\cdot x}\phi(|x-\tau^{-1}a|)$
is permutationally symmetric. Such families describe wave packets
with the momenta directed toward or away from the origin.
\end{remark} If for $d\geq 2$ the potential $V(x)$ is spherically
symmetric, then $V^{''}(0)=\frac{1}{d}\Delta V(0)\cdot Id_{n\times
n}$, and therefore the eigenvalues $e_{j}$ of $V^{''}(0)$ are all
equal to $ \frac{1}{d}\Delta V(0)$. Thus we have
\begin{corollary}\label{mainpo1}
Let $d\geq 2$ and $V$ satisfy Condition (VA) and let $V$ be
spherically symmetric. Then $L(\lambda)$ restricted to permutational
symmetric functions has $4$ eigenvectors or associated eigenvectors
near zero: two-dimensional space with eigenvalue 0; and
two-dimensional space with the non-zero imaginary eigenvalues $\pm
i\epsilon(\lambda)$, where $$\epsilon(\lambda)=h\sqrt{\frac{2\Delta
V(0)}{d}}+o(h),$$ and with the eigenfunctions
$\left(\begin{array}{lll}
\xi(\lambda)\\
\pm i\eta(\lambda)
\end{array}
\right)$, where $\xi$ and $\eta$ are real, and permutation symmetric
functions satisfying
$$\xi(\lambda)=\sqrt{2}\displaystyle\sum_{n=1}^{d}\frac{d}{dx_{n}}\phi_{0}^{\lambda}
+O(h)\ \text{and}\ \eta(\lambda)=-h\sqrt{\frac{1}{d}\Delta
V(0)}\displaystyle\sum_{n=1}^{d}x_{n}\phi^{\lambda}_{0} +O(h^{3/2})
.$$
\end{corollary}

The eigenvectors $\left(
\begin{array}{lll}
\xi(\lambda)\\
\pm i\eta(\lambda)
\end{array}
\right)$ are symmetric combinations of the eigenvectors described
in Theorem 2.

Besides eigenvalues, the operator $L(\lambda)$ may have resonances
at the tips, $\pm i\lambda$, of its essential spectrum (those tips
are called thresholds).
Recall the notation $\alpha_+ := \alpha$ if $\alpha>0$ and $=0$ of
$\alpha\leq 0$.
\begin{definition}
Let $d\neq 2$. A function $h$ is called a resonance function of
$L(\lambda)$ at $\mu=\pm i\lambda$ if $h\not\in \mathcal{L}^{2}$,
$|h(x)|\leq c\langle x\rangle^{-(d-2)_+}$ and $h$ is $C^{2}$ and
solves the equation
$$(L(\lambda)-\mu)h=0.$$
\end{definition}

Note that this definition implies that for $d>2$ the resonance
function $h$ solves the equation $(1+K(\lambda))h=0$ where
$K(\lambda)$ is a family of compact operators given by $K(\lambda)
:= (L_{0}(\lambda)-\mu+0)^{-1}V_{big}(\lambda)$. Here
$L_{0}(\lambda):=\left(
\begin{array}{lll}
0&-\Delta+\lambda\\
\Delta-\lambda&0
\end{array}
\right)$
and
\begin{equation}\label{Vbig}
V_{big}(\lambda):=\left(
\begin{array}{lll}
0&V_{h}-f((\phi^{\lambda})^{2})\\
-V_{h}+f((\phi^{\lambda})^{2})+2f^{'}((\phi^{\lambda})^{2})(\phi^{\lambda})^{2}&0
\end{array}
\right).
\end{equation}

In this paper we make the following assumptions on the point
spectrum and resonances of the operator $L(\lambda):$
\begin{enumerate}
 \item[(SA)] $L(\lambda)$ has only $4$ standard and associated
eigenvectors in the permutation symmetric subspace.
 \item[(SB)] $L(\lambda)$ has no resonances at $\pm i\lambda$.
\end{enumerate}

The discussion and results concerning these conditions, given in
~\cite{GS1}, suggested strongly that Condition (SA) is satisfied
for a large class of nonlinearities and potentials and Condition
(SB) is satisfied generically. Elsewhere
we show this using earlier
results of ~\cite{CP, CPV}. We also assume the following condition
\begin{enumerate}
 \item[(FGR)] Let $N$ be the smallest positive integer
such that $\epsilon(\lambda)(N+1)> \lambda,\ \forall \lambda \in
I$. Then $Re Y_{N}< 0$ where $Y_{n},\ n=1,2,\cdots,$ are the
functions of $V$ and $\lambda,$ defined in Equations
(~\ref{eq:tranform}) below (see also (~\ref{betadecay}).
\end{enumerate}

We expect that Condition (FGR) holds generically. Theorem
~\ref{maintheorem2} below shows that $Re Y_{n}= 0$ if $n<N.$

We expect the following is true:\\
(a) if for some $N_{1}(\geq N),$ $ReY_{n}=0$ for $n<N_{1},$ then
$ReY_{N_{1}}\leq 0$ and (b) for generic potentials/nonlinearities
there exists an $N_{1}(\geq N)$ such that $ReY_{N_{1}}\not=0.$ Thus
Condition (FGR) could have been generalized by assuming that $Re
Y_{N_1}< 0$ for some $N_1 \geq N$ such that $ReY_{n}=0$ for
$n<N_{1}$. We took $N=N_1$ in order not to complicate the
exposition.

The following form of $Re Y_{N}$
 \begin{equation}\label{FGR}
 Re Y_{N}=Im\langle \sigma_{1}(L(\lambda)-(N+1)i\epsilon(\lambda)-0)^{-1}F,F\rangle\leq 0
 \end{equation}
 for some function
 $F$ depending on $\lambda$ and $V$ and $\sigma_{1}:=\left(
 \begin{array}{lll}
 0&-1\\
 1&0
 \end{array}
 \right)$, is proved in ~\cite{BuSu, TY1, TY2, TY3, SW4} for $N=1,$ and in ~\cite{G} for $N=2,3$.
 We conjecture that this formula holds for any $N$.

 Condition (FGR) is related to the Fermi Golden Rule condition which
 appears whenever time-(quasi)periodic, spatially localized solutions
 become coupled to radiation. In the standard case it says that this coupling is effective
 in the second order ($N=1$) of the perturbation theory and therefore it leads to instability of such
 solutions. In our case these time-periodic solutions are
 stationary solutions
$$c_1 \left(
\begin{array}{lll}
\xi\\
 i\eta
\end{array}
\right)e^{i\epsilon(\lambda) t} +c_2 \left(
\begin{array}{lll}
\xi\\
- i\eta
\end{array}
\right)e^{-i\epsilon(\lambda) t}$$
 of the linearized equation $\frac{\partial\vec{\chi}}{\partial
 t}=L(\lambda)\vec{\chi}$ and the coupling is realized through the
 nonlinearity. Since the radiation in our case is "massive"$-$ the
 essential spectrum of $L(\lambda)$ has the gap
 $(-i\lambda,i\lambda)$, $\lambda>0,$ $-$ the coupling occurs only in
 the $N-$th order of perturbation theory where $N$ is given in
 Condition (FGR).

The rigorous form of the Fermi Golden Rule for the linear
Schr\"odinger  equations was introduced in ~\cite{BS}. For nonlinear
waves and Schr\"odinger equations the Fermi Golden Rule and the
corresponding condition were introduced in ~\cite{S} and, in the
present context, in ~\cite{SW4, BuSu, BP2, TY1, TY2, TY3}.

\section{Main Results}\label{sectionmaintheorem}
In this section we state the main theorem of this paper. For
technical reason we impose the following conditions on $f$ and $V$
\begin{enumerate}
 \item[(fC)] the nonlinearity $f$ is a smooth function satisfying $f^{''}(0)=f^{'''}(0)=0$
 if $d\geq 3$;
and $f^{(k)}(0)=0$ for $k=2,3\cdot\cdot\cdot 3N+1$ if $d=1,$ where
$f^{(k)}$ is the $k-$th derivative of $f$, and $N$ is the same as in
Condition (FGR),
 \item[(VB)] $V$ decays exponentially fast at $\infty.$
\end{enumerate}
\begin{theorem}\label{maintheorem1}
Let Conditions (fA)-(fC), (VA), (VB), (SA), (SB) and (FGR) be
satisfied and let, for $d\geq 3$, the potential $V$ be spherically
symmetric. Let an initial condition $\psi_0$ be permutation
symmetric if $d\geq 3$ and $\lambda\in \mathcal{I}$. There exists
$c,\epsilon_{0}>0$ such that, if
\begin{equation}\label{InitCond}\inf_{\gamma\in
\mathbb{R}}\{\|\psi_0-e^{i\gamma}(\phi^{\lambda}+z_{1}^{(0)}\xi+iz_{2}^{0}\eta)\|_{\mathcal{H}^{k}}+
\|(1+x^{2})^{\nu}[\psi_0-e^{i\gamma}(\phi^{\lambda}+z_{1}^{(0)}\xi+iz_{2}^{0}\eta)]\|_{2}\}\leq
c|(z_{1}^{0},z_{2}^{0})|^{2}
\end{equation} with
$|(z_{1}^{0},z_{2}^{0})|\leq \epsilon_{0}$ and $z_{n}^{0} \ n=1,2$
being real, some large constant $\nu>0$ and with $k=[\frac{d}{2}]+2$
if $d\geq 3,$ and $k=1$ if $d=1,$ then there exist differentiable
functions $\gamma,\ z_{1},\ z_{2}:\mathbb{R}^{+}\rightarrow
\mathbb{R},$ $\lambda:\mathbb{R}^{+}\rightarrow \mathcal{I}$ and
$R:\mathbb{R}^{+}\rightarrow \mathcal{H}^{k}$ such that the
solution, $\psi(t)$, to Equation (~\ref{NLS}) is of the form
\begin{equation}\label{Parametrization}
\psi(t)=e^{i \int_{0}^{t}
\lambda(s)ds+i\gamma(t)}[\phi^{\lambda(t)}+
z_{1}(t)\xi+iz_{2}(t)\eta+R(t)]
\end{equation} with the following
estimates:
\begin{enumerate}
 \item[(A)] $\displaystyle\|(1+x^{2})^{-\nu}R(t)\|_{2}\leq c(1+|t|)^{-\frac{1}{N}}$
 where $\nu$ and $N$ are the same as that in (~\ref{InitCond}) and
 (FGR) respectively,
 \item[(B)] $\displaystyle\sum_{j=1}^{2}|z_j(t)|\leq c(1+t)^{-\frac{1}{2N}}.$
\end{enumerate}
\end{theorem}
\begin{remark}
Recall from Remark ~\ref{remark2} that the class of permutationally
symmetric data includes wave packets with initial momenta directed
toward or in the opposite direction of the origin.
\end{remark}
\begin{theorem}\label{maintheorem2} Under the conditions of Theorem
3 we have
\begin{enumerate}
 \item[(A)] there exists a constant $\lambda_{\infty}\in \mathcal{I}$ such
 that $\displaystyle\lim_{t\rightarrow \infty}\lambda(t)=\lambda_{\infty}.$
 \item[(B)] Let
 $z:=z_{1}-iz_{2}$. Then there exists a change of variables
 $\beta=z+O(|z|^{2})$ such that
 \begin{equation}\label{betadecay}
 \dot{\beta}=i\epsilon(\lambda)\beta+\sum_{n=1}^{N}Y_{n}(\lambda)\beta^{n+1}\bar{\beta}^{n}+O(|\beta|^{2N+2})
 \end{equation}
\end{enumerate}
 with $Y_{n}$ being purely imaginary if $n<N$ and, by Condition (FGR) $Re
 Y_{N}<
 0$. Moreover,
 for $N=1,2,3$,  $ReY_{N}$ is given by Equation (~\ref{FGR}).

\end{theorem}
\begin{remark}
Using that $\epsilon(\lambda)=h\sqrt{\frac{2\Delta V(0)}{d}}+o(h)$
one can rewrite Equations (~\ref{Parametrization}) and
(~\ref{betadecay}) in the form (~\ref{decom1}) with $a(t)$ and
$p(t)$ satisfying the equations $\frac{1}{2}\dot{a}=p$ and
$\dot{p}=-h^{2}\nabla V(a)$ modulo $O(|a|^{2}+|p|^{2}).$
\end{remark}
\section{Idea of the Proof: $z$-Expansions}\label{secasymptotic}
We follow ~\cite{G, GS2}. We decompose of the solution $\psi(t)$ to
Equation (~\ref{NLS}) into a solitonic component and a
simplectically orthogonal fluctuation as (cf ~\cite{FGJS})
\begin{equation}\label{decompositionR}
\psi(t)=e^{i \int_{0}^{t}
\lambda(s)ds+i\gamma(t)}(\phi^{\lambda}+z_{1}(t)\xi+iz_{2}(t)\eta+R(t)),
\end{equation}
where $\lambda,\ \gamma,\ z_1,\ z_2 $ are real, differentiable
functions of $t$ and the function $R(t)$, called the
\textit{fluctuation}, satisfies the orthogonality conditions
\begin{equation}\label{Rorthogonal}
Im\langle R,i\phi^{\lambda}\rangle=Im\langle
R,\frac{d}{d\lambda}\phi^{\lambda}\rangle=Im\langle
R,i\eta\rangle=Im\langle R,\xi\rangle=0.
\end{equation}

We plug Equation (~\ref{decompositionR}) into Equation
(~\ref{NLS}) to obtain equations for the parameters $z_{1}(t),
z_{2}(t), \lambda(t)$ and $\gamma(t)$ and the fluctuation $R(t)$.
As was already discussed above the linearized operator,
$G(\phi^{\lambda})$ (see (~\ref{defineoperator})), in the equation
for $R(t)$ is only real-linear and therefore we pass from the
unknown $R$ to the unknown $\vec{R}:=\left(
\begin{array}{lll}
Re R\\
Im R
\end{array}
\right)\leftrightarrow R$. Under this correspondence the
multiplication by $i^{-1}$ goes over to the symplectic matrix
$J:=\left(
\begin{array}{lll}
0&1\\
-1&0
\end{array}
\right):\ J\vec{R}\leftrightarrow i^{-1}R.$   Unlike with the
equation for $R$, in the equations for $z_{1}$ and $z_{2}$ it is
more convenient to go from the real, symplectic structure given by
$J$ to the complex structure $i^{-1}$ by passing from $\left(
\begin{array}{lll}
z_{1}\\
z_{2}
\end{array}
\right)$ to $z:=z_{1}-iz_{2}$.

A key point is to look for the fluctuation $\vec{R}(t)$ in the form
(~\cite{G, GS2})
\begin{equation}\label{eq:decom}
\vec{R}=\sum_{2\leq m+n\leq
N}\tilde{R}_{m,n}(\lambda)z^{m}\bar{z}^{n}+\tilde{R}_{N}
\end{equation}
with the remainder $\tilde{R}_{N}$ of the order $O(|z|^{N+1})$.
This leads to the following equations on the coefficients
$\tilde{R}_{mn}$:

$$[L(\lambda)-i\epsilon(\lambda)(m-n)]\tilde{R}_{mn}(\lambda)=-P_{c}\tilde{f}_{m,n}(\lambda),$$
where the functions $\tilde{f}_{m,n}(\lambda)$ depend on
$\tilde{R}_{m',n'}(\lambda)$  with $m'+n'<m+n$. Recall that if
$|m-n|\leq N,$ then $i\epsilon(\lambda)(m-n)\not\in
\sigma(L(\lambda))$ and therefore the operators
\begin{equation}\label{Invers}
L(\lambda)-i\epsilon(\lambda)(m-n):P_{c}\mathcal{L}^{2}\rightarrow
P_{c}\mathcal{L}^{2}
\end{equation} are invertible. Hence  the above equations
have unique solutions.

We plug the expansion (~\ref{eq:decom})
the differential equations for the parameters $\lambda,\ \gamma$
and $z$ to obtain expansions for $\dot\lambda,\ \dot\gamma$ and
$\dot{z}$ in terms of $z$ and $\bar{z}$.

In the second step we transform $z$ to a parameter $y$ which
satisfies a simpler different equation. We construct a polynomial
$P(z,\bar{z})$ with real coefficients and the smallest degree $\geq
2$, such that the new parameter $y:=z+P(z,\bar{z})$ satisfies the
equation
\begin{equation}\label{eq:tranform}
\begin{array}{lll}
\dot{y}&=&i\epsilon(\lambda)y+\displaystyle\sum_{2\leq m+n\leq
2N+1}Y_{m,n}(\lambda)y^{m}\bar{y}^{n}\\& &+Remainder.
\end{array}
\end{equation} where the coefficients $Y_{m,n}(\lambda)$
are purely imaginary, and $Y_{m,n}(\lambda)=0$ if $m\neq n+1$, and
the term $Remainder$ admits the estimate $$|Remainder|\leq
ct^{-\frac{2N+1}{2N}}\ \text{or}\ |y(t)|^{2N+1}+|y(t)|^{N}\|\langle
x\rangle^{-\mu}R_{N}\|_{2}+\|\langle
x\rangle^{-\mu}R_{N}\|^{2}_{2}.$$

Expressing the variables $z$ and $\bar{z}$ as power series in $y$
and $\bar{y}$ and plugging the result into the $z-$expansions for
$\dot\gamma$, $\dot\lambda$ and $\vec{R}$ mentioned above
we obtain the $y-$expansions for these quantities.

As was mentioned above, the decrease of $z$ (or $y$) and therefore
the relaxation of the soliton to its equilibrium occurs due to the
radiation of the excess of the energy to infinity. The latter is
possible only if the periodic solutions to the linearized equation
are coupled to its continuous spectrum. To detect this coupling
we must obtain an expansion of $\vec{R}$ in the parameters $y$ and
$\bar{y}$ up to the order $2N$.
This is done in the third, most involved step. As in the first
step we determine the coefficients $R_{m,n}(\lambda)$ of the
$y-$expansion of $\vec{R}$ by solving the equations
$$[L(\lambda)-i\epsilon(\lambda)(m-n)]R_{m,n}(\lambda)=-P_{c}f_{m,n}(\lambda)$$ for certain functions $f_{m,n}(\lambda)$ (see below). Recall that the number $N$ is defined by the properties
$$i\epsilon(\lambda)(m-n)
\begin{array}{lll}
\not\in\sigma(L(\lambda))\ \text{if}\ |m-n|\leq N,\\
\in\sigma(L(\lambda))\ \text{if}\ |m-n|> N.
\end{array}
$$
Thus for $N<m+n\leq 2N$ the parameter $i\epsilon(\lambda)(m-n)$
might be in the spectrum of $L(\lambda).$ To deal with this case
we sort out the pairs $(m,n)$ into "non-resonant pairs" satisfying
$|m-n|\leq N$ and "resonant pairs" satisfying $|m-n|> N.$ For
"non-resonant" pairs the operators
$$L(\lambda)-i\epsilon(\lambda)(m-n): P_{c}\mathcal{L}^{2}\rightarrow
P_{c}\mathcal{L}^{2}$$ are invertible and for resonant pairs they
are not (one has to change spaces in the latter case).

In the first two steps we expanded in $z$ and $\bar{z}$ (and in
$y$ and $\bar{y}$) until $m+n\leq N$ and consequently all the
pairs, $(m,n)$, involved were non-resonant ones. Now our expansion
involves pairs $(m,n)$ with $m+n>N,$ which include resonant pairs.
A key point which we show is that for the subsets of pairs
$(m,n)$, $m+n>N,$ determined by the inequality
$$m,n\leq N,$$ the terms $f_{m,n}(\lambda)$ involve only "non-resonant" pairs (here
we use that the parameter $y$ satisfies (~\ref{eq:tranform}) with
$Y_{m,n}(\lambda)=0$ for $m\neq n+1$) and consequently we are able
to solve for the coefficients $R_{m,n}(\lambda)$ also in this
case.

Finally, estimates of the remainders as in Equations
(~\ref{eq:decom})  are done by rewriting the differential equations
for them in an integral form (using the Duhamel principle) and using
estimates of linear propagators derived in
~\cite{Buslaev,BP2,BuSu,RSS,GoSc,GS1,GS2}.

\end{document}